\documentclass[pdflatex,sn-mathphys-num]{sn-jnl}
\usepackage{float}
\usepackage{anyfontsize}
\usepackage{graphicx}%
\usepackage{multirow}%
\usepackage{amsmath,amssymb,amsfonts}%
\usepackage{amsthm}%
\usepackage{mathrsfs}%
\usepackage[title]{appendix}%
\usepackage{xcolor}%
\usepackage{textcomp}%
\usepackage{manyfoot}%
\usepackage{booktabs}%
\usepackage{algorithm}%
\usepackage{algorithmicx}%
\usepackage{algpseudocode}%
\usepackage{listings}%
\usepackage{minted}
\usepackage{csquotes}
\usepackage{siunitx}
\usepackage{braket}
\definecolor{codebg}{RGB}{248,248,248}
\definecolor{codeadd}{RGB}{220,255,220} 
\definecolor{coderm}{RGB}{255,220,220}  
\theoremstyle{thmstyleone}%
\usepackage{tabularx}
\usepackage{makecell}
\usepackage{xurl} 

\theoremstyle{thmstyletwo}%

\theoremstyle{thmstylethree}%
\DeclareSIUnit\hartree{Ha}
\DeclareSIUnit\angstrom{\text {Å}}
\begin{document}

\title{Automated near-term quantum algorithm discovery for molecular ground states} 

\author{\centering
Fabian Finger$^\ddagger$, Frederic Rapp$^\ddagger$, Pranav Kalidindi$^\ddagger$,\\
Alexander Koziell-Pipe, David Zsolt Manrique, Gabriel Greene-Diniz,\\
Stephen Clark, Konstantinos Meichanetzidis$^\S$\\
\vspace{3mm}
\emph{Quantinuum}\\
\vspace{8mm}
Kerry He$^\S$, Kante Yin, Hamza Fawzi,\\
Bernardino Romera Paredes, Alhussein Fawzi\\
\vspace{3mm}
\emph{Hiverge}
\footnote[0]{$\ddagger$: equal contribution}
\footnote[0]{$\S$: Correspondence to: k.mei@quantinuum.com and kerry@hiverge.ai}
}



\abstract{
Designing quantum algorithms is a complex and counterintuitive task,
making it an ideal candidate for AI-driven algorithm discovery.
To this end, we employ the \emph{Hive}, an AI platform for program synthesis,
which utilises large language models to drive a highly distributed evolutionary process for discovering new algorithms.
We focus on the ground state problem in quantum chemistry, and discover efficient quantum heuristic algorithms that solve it for molecules LiH, H$_2$O, and F$_2$ while exhibiting significant reductions in quantum resources relative to state-of-the-art near-term quantum algorithms.
Further, we perform an interpretability study on the discovered algorithms and identify the key functions responsible for the efficiency gains.
Finally, we benchmark the \emph{Hive}-discovered circuits on the Quantinuum System Model H2 quantum computer and identify minimum system requirements for chemical precision.
We envision that this novel approach to quantum algorithm discovery applies to other domains beyond chemistry, as well as to designing quantum algorithms for fault-tolerant quantum computers.
}
\maketitle

\section{Introduction}
Quantum computing and machine learning are two rapidly developing fields that are reshaping modern computational science. 
Quantum computing has the potential to outperform classical systems on specific classes of problems, but achieving this hinges on the development of efficient quantum algorithms~\cite{feynman_simulatingphysics_1982}. 
However, designing quantum algorithms is conceptually challenging as one needs to manipulate quantum phenomena such as interference, entanglement, and superposition. 
Furthermore, quantum hardware is limited in terms of the number of qubits and gate fidelities due to inherent noise and the required overheads of quantum error correction (QEC) \cite{preskill_quantumcomputing_2018,strohm_ionbasedquantum_2024}.
This constrains the types of quantum algorithms that can be executed, which are beyond the reach of classical algorithms.
Therefore, optimising the quantum resources required by quantum algorithms is essential for bridging the gap between current noisy hardware and practically useful quantum computers.
In parallel, AI systems have begun to show great promise in discovering and optimising algorithms autonomously~\cite{AlphaTensor2022,romera-paredes_mathematicaldiscoveries_2024, lange_shinkaevolveopenended_2025, sharma_openevolveopensource_2025, surina_algorithmdiscovery_2025, novikov_alphaevolvecoding_2025}.
Evolutionary style approaches, in combination with large language models, have produced novel, human-understandable algorithms that have performed better than known human-designed algorithms~\cite{novikov_alphaevolvecoding_2025}. 
These AI-driven algorithm discovery frameworks optimise \enquote{how to solve} rather than merely \enquote{the solution}, and have discovered heuristics that remarkably surpass the state of the art for hard
computational problems~\cite{romera-paredes_mathematicaldiscoveries_2024,novikov_alphaevolvecoding_2025,lange_shinkaevolveopenended_2025}.

AI-driven algorithm discovery promises to optimise existing algorithms or even discover entirely novel quantum algorithms across a wide range of application domains \cite{ruiz_quantumcircuit_2025,acampora_quantumcomputing_2025,alexeev_artificialintelligence_2025, krenn_artificialintelligence_2023,knipfer_aiagents_2026}.
In this work, we focus on quantum chemistry, a leading long-term application area for quantum computing.
At the heart of quantum chemistry lies the electronic structure problem: determining molecular ground-state energies to chemical precision, that is, to the level required to predict chemically relevant quantities such as reaction rates and binding energies~\cite{aspuru-guzik_simulatedquantum_2005,cao_quantumchemistry_2019,mcardle_quantumcomputational_2020}.

The relevant quantum algorithm is Quantum Phase Estimation (QPE)~\cite{kitaev_quantummeasurements_1995, nielsen_quantumcomputation_2010}, which, under the assumption of an initial state with high enough overlap with the ground state, is expected to prepare the ground state in polynomial time.
Early proof-of-concept implementations of QPE on quantum hardware using QEC protocols have already been demonstrated for the H$_2$ molecule~\cite{yamamoto_quantumerrorcorrected_2025}.
However, resource requirements remain substantial for anything beyond the smallest of molecules.
Thus, we turn to variational quantum algorithms whose resource requirements, both for classical simulation and for implementation on near-term quantum processors, are lower.
In particular, we benchmark two adaptive protocols within the variational quantum eigensolver (VQE) framework~\cite{peruzzo_variationaleigenvalue_2014}: the ADAPT-VQE~\cite{grimsley_adaptivevariational_2019} and the qubit-excitation-based (QEB)-ADAPT-VQE~\cite{yordanov_qubitexcitationbasedadaptive_2021} algorithms.
These algorithms are not expected to scale to large molecules, as, in the worst case, they suffer from barren plateaus that hinder convergence~\cite{larocca_barrenplateaus_2025} and require unrealistically high numbers of shots to achieve chemical precision.
Nevertheless, this class of quantum heuristics offers a rich playground for exploring and developing novel AI-driven techniques for quantum algorithm discovery.

To this end, we leverage the \emph{Hive}, an AI platform for program synthesis, and demonstrate that it can autonomously discover efficient VQE-style algorithms to find the ground states of molecules such as Lithium Hydride (LiH), Water (H$_2$O), and Fluorine (F$_2$).
Even though the ground states for these molecules are known via exact diagonalisation, solving them using near-term methods can push the state of the art of VQE algorithms to their limits.
We demonstrate how the discovered VQE variants achieve chemical precision while showing significant reductions in quantum resources, such as two-qubit gates and circuit evaluations, compared with well-known adaptive VQE-style algorithms.
In addition, we investigate the generalisation ability of the discovered quantum heuristic to other bond lengths for a given molecule, and perform an interpretability study to explain the source of the performance gains.
Finally, we benchmark Quantinuum's System Model H2-1 quantum processor using the circuits output by one of the \emph{Hive}-discovered algorithms and predict noise parameters such that chemical precision is retained.

\section{AI for Algorithm Discovery and Related Work}
\label{sec: AI for QA}
As the capabilities of large language models (LLMs) continue to advance, there is growing interest in using these models for scientific research, such as designing novel state-of-the-art algorithms. 
However, while code generation is a primary application of LLMs, these models are inherently prone to hallucinations \cite{huang_surveyhallucination_2025,zhang_llmhallucinations_2025}. 
Without proper safeguards, there is no guarantee that the code generated by the LLM is functionally correct. 
To address these issues, AI algorithm discovery frameworks introduce scaffolds around LLMs to improve their performance and guarantee correctness. 
The key idea is to explicitly run and evaluate each program output by the LLM, then allow the LLM to iteratively improve and refine upon its past outputs using this grounded feedback. 
These methods were first popularised by FunSearch~\cite{romera-paredes_mathematicaldiscoveries_2024} and its successor AlphaEvolve~\cite{novikov_alphaevolvecoding_2025}, and there have since been multiple variations and improvements on these systems~\cite{surina_algorithmdiscovery_2025,lange_shinkaevolveopenended_2025,wan_loongflowdirected_2025}. 

\begin{figure}[t]
    \centering
    \includegraphics[width=\linewidth]{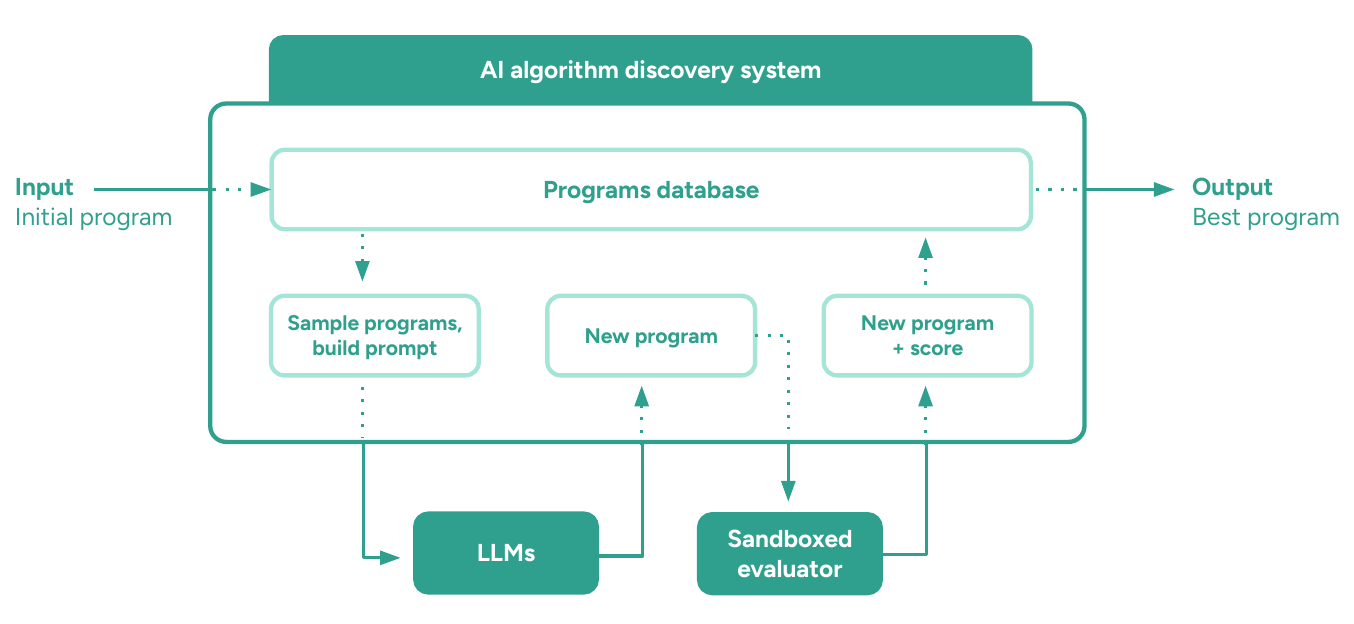}
    \caption{\textbf{Overview of AI algorithm discovery frameworks.} The system samples high-performing programs from a programs database that serve as the basis for improvements proposed by an LLM. New programs are evaluated in a sandbox to generate a quantifiable performance score, and both the program and its score are stored in the programs database to use in future iterations.
    }
    \label{fig:EvoAI4Algo}
\end{figure}

The general workflow of AI algorithm discovery frameworks is illustrated in Figure~\ref{fig:EvoAI4Algo}.
As input into the system, a human user specifies the code that the system is tasked to evolve, and provides an evaluator that validates the correctness of the program and quantifies how well the program performs at the given task. 
Starting from this initial user-specified code, the system prompts an LLM to suggest improvements to the code, either by outputting code directly or outputting modifications to apply as patches to the code. 
The resulting program is then evaluated in a sandboxed environment, given a quantifiable score, and stored in a programs database if it passes the correctness checks.
This loop then continues using search heuristics, such as evolutionary algorithms or Monte Carlo tree search. 
At a high level, high-performing programs in the database are chosen for LLMs to further suggest improvements to the code. 
Information from other programs can also be used to aid the LLM prompt, such as by providing another high-performing program for the LLM to draw ideas from, or by providing a summary of the state of the programs database to inform the LLM about the landscape of programs it has already explored.
Upon termination, the AI algorithm discovery agent then outputs the program it found with the best evaluated metric.

Notably, these methods have found success in various mathematical and scientific applications~\cite{romera-paredes_mathematicaldiscoveries_2024,novikov_alphaevolvecoding_2025,georgiev_mathematicalexploration_2025,zhang_quantumcomputation_2025}. 
We remark that many scientific algorithms contain heuristic, handcrafted components that lack formal guarantees and are instead refined through manual experimentation and expert intuition. 
These heuristics are particularly suited for AI-driven algorithm discovery systems, which can efficiently explore these vast algorithmic spaces. 
Additionally, unlike classical reinforcement-learning approaches, which output a policy network, because all candidate algorithms found by LLM-driven methods are explicit programs, the discovered algorithms remain interpretable, can leverage established software libraries, and are amenable to inspection and analysis by domain experts, or even LLMs.

In this work, we use Hiverge's proprietary implementation of these AI algorithm discovery agents, which we refer to as the \emph{Hive}.
Notably, this system supports the evolution of codebases consisting of multiple files and directories, runs in a distributed fashion, and supports sandboxes consisting of cloud GPUs. 

\section{The Electronic Structure Problem}
In this work, we focus on the ground-state search problem in quantum chemistry  \cite{aspuru-guzik_simulatedquantum_2005,cao_quantumchemistry_2019,mcardle_quantumcomputational_2020}. 
Given a many-electron Hamiltonian describing electron-electron interactions in a fixed nuclear potential, the task is to determine the Hamiltonian's lowest energy
eigenstate and the corresponding eigenvalue. 
An accurate description of the molecular ground state is essential for predicting equilibrium structures, chemical properties, and reactivity, which are crucial for various applications in drug discovery and material science \cite{reiher_elucidatingreaction_2017, cao_potentialquantum_2018, babbush_lowdepthquantum_2018}.

However, obtaining this ground state is computationally challenging. Classical algorithms suffer from unfavourable scaling when solving the problem exactly. Approximate methods such as Hartree--Fock, Density Functional Theory (DFT) \cite{hohenberg_inhomogeneouselectron_1964, kohn_selfconsistentequations_1965}, and Coupled-Cluster \cite{helgaker_molecularelectronicstructure_2013} rely on approximations that trade precision for tractability and may fail in strongly correlated systems \cite{cooper_benchmarkstudies_2010,vanvoorhis_benchmarkvariational_2000, evangelista_alternativesinglereference_2011}. As a result, there is significant interest in exploring quantum algorithms that could potentially overcome these limitations.

\subsection{Near-Term Algorithms}
\begin{figure}[t]
    \centering
    \includegraphics[width=\linewidth]{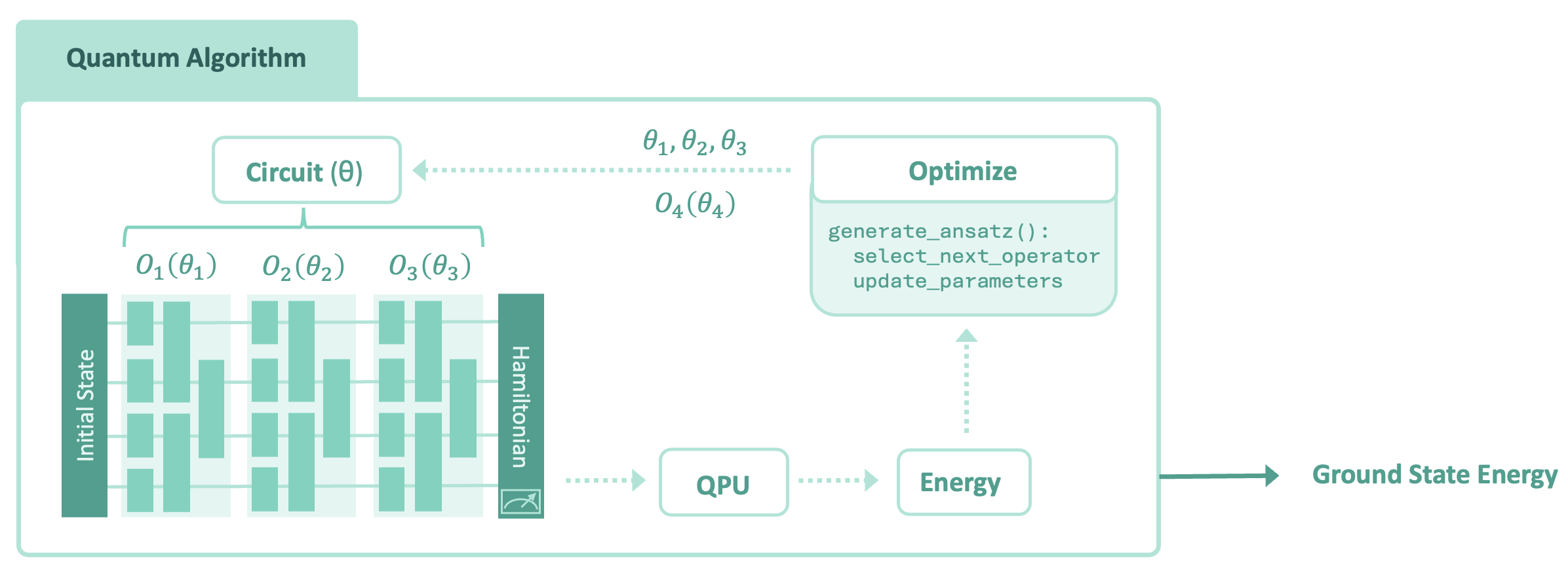}
    \caption{\textbf{VQE-style quantum algorithms.} This family of quantum algorithms is defined by two components: a parametrised quantum circuit and a classical optimisation step. The aim is to find the ground state of a molecule, so we seek a trial state that yields the lowest eigenvalue of the molecule's Hamiltonian. We first define the Hamiltonian of a molecule in a chosen basis and at a given geometry, the initial state, typically the Hartree–Fock state, and an operator pool $\{O_i(\theta_i)\}$. These parameterised operators are used by the classical optimisation step to generate an ansatz and update its parameters. The trial state is evaluated, and based on the results, the classical step updates the ansatz and its parameters iteratively in a feedback loop.}
    \label{fig:vqeskeleton}
\end{figure}

Variational Quantum Algorithms (VQAs) \cite{cerezo_variationalquantum_2021} are a family of hybrid quantum-classical algorithms that are well suited for noisy intermediate-scale quantum (NISQ) devices. 
They use a parameterised quantum circuit (ansatz) to prepare a trial state, with parameters optimised through a feedback loop between the quantum device and a classical optimiser to minimise (or maximise) a chosen objective function. 
Within this family of algorithms, we focus on the VQE~\cite{peruzzo_variationaleigenvalue_2014}, and its ADAPT-VQE variants~\cite{grimsley_adaptivevariational_2019, yordanov_qubitexcitationbasedadaptive_2021}, which estimate the lowest eigenvalue of a given Hamiltonian and are among the most practical near-term approaches for molecular ground-state calculations. 
While VQE optimises a fixed ansatz, ADAPT-VQE iteratively constructs the ansatz by selecting operators from a predefined operator pool based on their energy gradients, yielding more compact and expressive circuits. 
Figure \ref{fig:vqeskeleton} shows the workflow for applying VQE-style algorithms to find the ground state of a specific molecule.

A widely used chemistry-inspired ansatz is the Unitary Coupled Cluster with Singles and Doubles (UCCSD) ansatz~\cite{anand_quantumcomputing_2022}, which comprises single and double excitation operators that act on the spin orbitals of the molecule. 
Alternatives such as UCC-inspired qubit excitation pools~\cite{yordanov_efficientquantum_2020} are utilised in this work as they produce more compact, hardware-efficient circuits to represent the excitations \cite{arrazola_universalquantum_2022} (see Appendix~\ref{app:qubit_excitations}).
Once the ansatz of operators $\{O_i(\theta_i)\}$ is applied to the initial state, the quantum device evaluates the energy. 
The classical optimiser uses the energy to generate a new ansatz to optimise the parameters $\{\theta_i\}$ and, in the ADAPT-VQE case, may also update the operators $\{O_i\}$ selected from the pool. 
This loop continues until convergence of the energy objective.

VQE is highly costly to run on NISQ devices, primarily due to the number of circuit evaluations.
Given that one aims to execute such algorithms on current hardware, it is important to minimise quantum resource requirements. 
In addition, the noise in these devices limits the depth of circuits that can be reliably executed. 
Although ADAPT-VQE mitigates this by constructing a more compact ansatz, it incurs its own overhead in the form of repeated gradient and Hamiltonian evaluations on the quantum device.
To discover more efficient VQE variants, we turn to the \emph{Hive} to minimise both the number of circuit evaluations and the circuit depth.

\section{Automated VQE-discovery with the \emph{Hive}}
\begin{figure}[t]
    \centering
    \includegraphics[width=\linewidth]{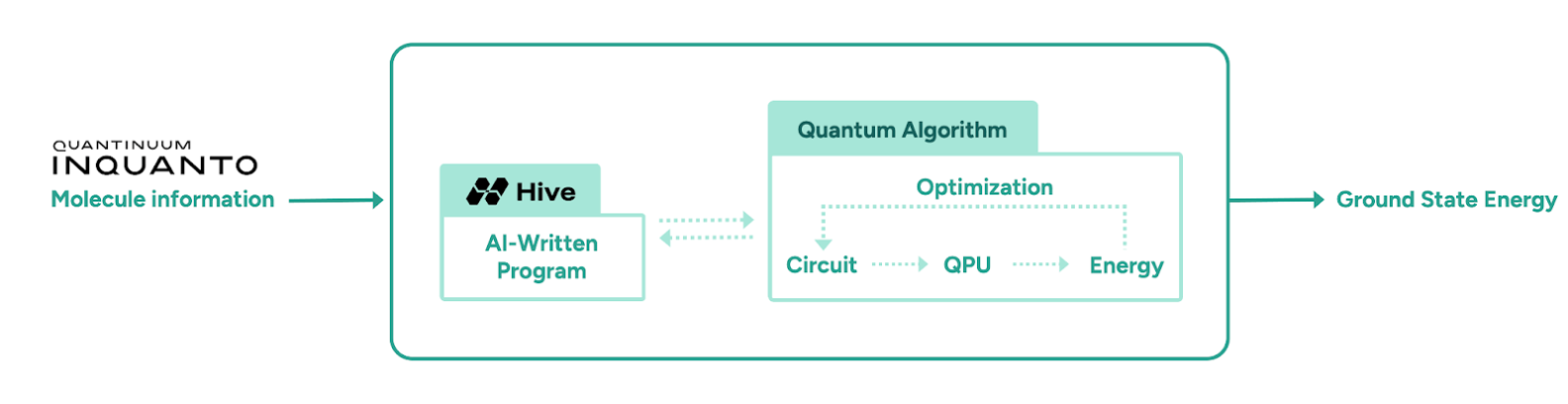}
    \caption{\textbf{Quantum algorithm discovery workflow.} An application of the workflow shown in Fig.~\ref{fig:EvoAI4Algo} to the quantum algorithm skeleton shown in Fig.~\ref{fig:vqeskeleton}. The \emph{Hive} is provided with a problem skeleton and information about the molecule at hand, which contains basic code to construct and evaluate quantum operators implemented by \emph{InQuanto}. It is also given a generic prompt that contains a basic description of the molecular ground-state problem in quantum chemistry. The candidate algorithms are evaluated on a backend, a quantum computer, or a classical simulator or emulator, which returns the energy value that serves as the main objective for the \emph{Hive} to optimise.
    }
    \label{fig:workflow}
\end{figure}

We task the \emph{Hive} with autonomously designing quantum programs that improve the convergence efficiency of VQE-style algorithms.
Specifically, the functions to be evolved are those contained in the \enquote{Optimise} part of the VQE workflow shown in Figure \ref{fig:vqeskeleton}, which define heuristic strategies for creating and optimising a quantum circuit ansatz, and the resulting discovery loop is shown in Figure \ref{fig:workflow}.
The \emph{Hive} is given a program skeleton, which provides the fundamental infrastructure for circuit construction, operator placement, and energy evaluation.

The \emph{Hive} has the freedom to import and use any package and library required.
Through Quantinuum's \emph{InQuanto}~\cite{inquanto}, a quantum computational chemistry library, the system can access molecular information such as the atomic geometry of a given molecule, basis set, and the Hamiltonian. 
\emph{InQuanto} offers predefined operator pools such as the UCCSD pool, its fully generalised variant UCCGSD \cite{bartlett_coupledclustertheory_2007, anand_quantumcomputing_2022, lee_generalizedunitary_2019}, and the one-layer $k$-UpCCGSD variant with generalised singles and pair doubles \cite{lee_generalizedunitary_2019}.
We utilise \emph{InQuanto} to obtain the relevant orbital combinations that these operator pools act on and NVIDIA's \textit{CUDA-Q} library~\cite{kim_cudaquantum_2025} to directly implement the corresponding qubit excitation pool as shown in Refs. \cite{arrazola_universalquantum_2022, yordanov_efficientquantum_2020} to produce hardware-efficient excitation operators (see Appendix \ref{app:qubit_excitations}).
These components define the computational environment within which new routines can be tested.
NVIDIA's \textit{CUDA-Q} library allows us to leverage GPUs for quantum circuit simulation, which is compatible with the \emph{Hive}'s general evaluation workflow, and offers the opportunity to distribute workloads. 

The main component of the program skeleton, and the function the \emph{Hive} is tasked to evolve, is the \texttt{generate\_ansatz()} function.
This function depends on the given molecular Hamiltonian (parameterised by bond length and basis set), and aims to assemble a quantum circuit from the available operator pools to estimate the ground state energy. 
The initial implementation includes several helper functions, e.g., for scoring excitations, deciding termination conditions, and selecting rotation angles. In the beginning, these helper functions use trivial and naive logic to avoid biasing the \emph{Hive} towards exploring any specific direction.
The evaluator given to the \emph{Hive} uses the \texttt{generate\_ansatz()} function to produce an ansatz for a given molecule and bond length, applies this to a reference state, and then evaluates the energy of the output state.
The evaluator also checks if the produced ansatz is physically valid, e.g., by checking if the largest qubit number used in the operators is less than the total number of qubits.
We also use the \emph{Hive} to minimise resource overhead by first evolving routines that achieve chemical precision and then retuning the fitness function to favour fewer circuit evaluations and lower two-qubit gate counts while preserving that accuracy target.

Beyond the code skeleton, \emph{Hive} receives structured natural language prompts that contextualise the optimisation goals and guide acceptable search directions. These prompts summarise essential quantum chemistry concepts (such as the electronic structure problem) and encode practical constraints, e.g., avoiding gradient evaluations, favouring compact circuit structures, and seeking low-energy solutions through adaptive or heuristic approaches. 
To illustrate the high-level constraints and heuristic hints exposed to the \emph{Hive} during search, Table~\ref{tab:quantum_chem_prompt_summary} shows one representative prompt layout.
We provide the skeleton in a GitHub repository\footnote{https://github.com/Quantinuum/quantum-algo-discovery-for-molecular-ground-states}.

Overall, this framework allows the \emph{Hive} to explore a combined algorithmic space of both variational and heuristic quantum procedures. By evolving only the algorithmic core within a chemically consistent environment, the system can autonomously propose approaches to ground-state estimation under NISQ-era conditions.


\begin{table}[ht!]
\centering
\setlength{\tabcolsep}{8pt}
\begin{tabular}{|p{\linewidth}|}
\hline
\textbf{General Prompt} \\ 
\hline
You are an expert in NISQ quantum computation, variational/heuristic quantum algorithms, and quantum chemistry. 
The project focuses on determining the ground state energy of H$_2$O in STO-3G using a UCCSD operator pool (single and double excitations). 
Your objective is to achieve \textit{chemical precision} (1.6 mHa) over bond lengths from 1.0~\AA{} to 3.0~\AA{}. 
Use \emph{InQuanto} for molecular data, CUDA-Q for quantum circuit execution, and OpenFermion for electronic structure information. 

Backends are expensive: \textbf{any approach that increases backend calls will be rejected}. 
The main scoring metric is how well the solution reaches chemical precision across all bond lengths (via \texttt{ChemistryDriverPySCFMolecularROHF} with \texttt{point\_group\_symmetry=True}). 

General implementation hints include caching, classical pre-filtering (\texttt{score\_excitation}), parametrised kernels, deterministic reproducibility, efficient ansatz construction, and adaptive optimisation near convergence.
Reuse the provided \texttt{\_evaluate\_ansatz\_energy} function, which manages caching and asynchronous execution. \\
\hline
\end{tabular}

\vspace{.5em}

\begin{tabular}{|p{0.4\linewidth}|p{0.55\linewidth}|}
\hline
\textbf{Idea / Technique} & \textbf{Description} \\
\hline
Code Simplification & Combine and clean subroutines; remove redundant code while maintaining precision and performance. \\ 
\hline
Fractional Occupancy Scoring & Use orbital occupations from OpenFermion to boost excitations with partially filled (active) orbitals, improving correlation handling. \\ 
\hline
Coordinate Descent with Freezing & Optimise only the newest parameter upon adding an operator; re-optimise all angles periodically to save backend calls. \\ 
\hline
Spatial Symmetry Filtering & Enforce orbital symmetry conservation; discard excitations breaking the molecule's spatial symmetry to shrink
the search space. \\ 
\hline
Frozen Front Optimisation & Optimise only the latest operator each iteration; perform global re-optimisation every $K$ steps or under stagnation, cutting backend calls by $\sim$80\%. \\ 
\hline
Look-Ahead Screening & Test top-$5$ classical candidates with a one-shot backend energy probe ($\theta=0.1$) to pick the best promising excitation. \\  
\hline
Entanglement-Weighted Scoring & Penalise non-local excitations (long-range qubits) unless they have a very high classical score; promotes hardware-efficient, entanglement-sensitive ansatz generation. \\ 
\hline
Symmetry-Enforced Pruning & Explicitly forbid excitations violating orbital irreducible representations ($A\!\to\!B$ if $A\neq B$), halving the search space and improving convergence. \\
\hline
\end{tabular}
\caption{Example layout of a structured natural language prompt for a quantum chemistry optimisation workflow, demonstrated on the H$_2$O molecule in the STO-3G basis using a qubit excitation operator pool. The Table summarises the hierarchical prompt composition: first defining the global objective (chemical-precision ground state energy estimation) and implementation constraints, followed by incremental heuristic strategies enhancing efficiency and variational performance in NISQ-era simulations.
}
\label{tab:quantum_chem_prompt_summary}
\end{table}

\section{Results}
In this Section, we benchmark the \emph{Hive}-discovered VQE-style algorithms against ADAPT-VQE and QEB-ADAPT-VQE along the LiH, H$_2$O, and F$_2$ dissociation curves, spanning bond-length scans from near-equilibrium to stretched geometries.
We show that the evolved routines sustain chemical precision on unseen bond lengths while substantially reducing circuit evaluations and two-qubit-gate counts, and we attribute these gains to a small set of interpretable mechanisms that remain effective under realistic noise models of the Quantinuum H2-1 emulator and hardware.

\subsection{Chemical Precision and Bond Length Generalisation}
We evaluate the algorithms discovered by the \emph{Hive} across the dissociation curves of three molecules selected to probe increasingly complex systems: LiH, H$_2$O, and F$_2$ in the STO-3G basis set. 
We benchmark against two canonical adaptive baselines implemented in our codebase: ADAPT-VQE~\cite{grimsley_adaptivevariational_2019}, which ranks operators by gradient magnitude and appends the top candidate, and its QEB variant~\cite{yordanov_qubitexcitationbasedadaptive_2021} which optimises the top $k=10$ gradient-ranked candidates via additional VQE steps and appends the operator with the largest realised energy decrease, trading a constant-factor increase in circuit evaluations for more reliable per-step progress (see Appendix~\ref{app:baselines}).

These benchmarks utilise the full orbital space without active space reduction, resulting in system sizes of 12, 14, and 20 qubits, respectively. 
We initialise all methods from the Hartree--Fock single-reference state $|\Phi_{\text{HF}}\rangle$, which poses a significant challenge in the dissociation regime, where the ground state exhibits strong multi-reference character that is difficult to capture from a mean-field starting point~\cite{lyakh_multireferencenature_2012}.

\begin{figure}[t]
    \centering
    \includegraphics{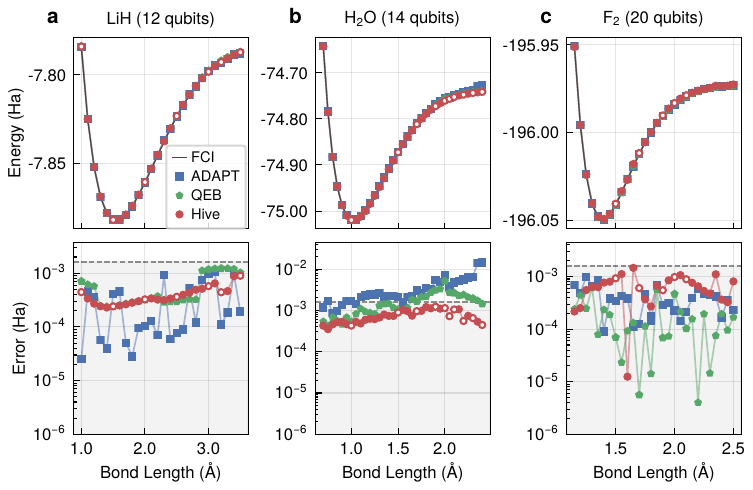}
    \caption{\textbf{Energy precision of bond dissociation curves obtained with the \emph{Hive} algorithms and ADAPT baselines as a function of bond length.} \textbf{a}--\textbf{c}, Bond dissociation curves (top panels) and energy errors relative to FCI (bottom panels) for LiH (\textbf{a}), H$_2$O (\textbf{b}), and F$_2$ (\textbf{c}). \emph{Hive} results are shown as red circles, ADAPT-VQE and QEB-ADAPT-VQE are shown as blue squares and green pentagons, respectively. Hollow markers indicate the specific bond lengths used in the evolution of the \emph{Hive} ansatz, while solid markers represent test points evaluated using the evolved algorithms. The dashed grey line in the error plots marks the chemical precision threshold (\qty{1.6e-3}{\hartree}). Bond length for LiH denotes the Li--H distance; for H$_2$O, the O--H bond length (with both O--H bonds stretched symmetrically); and for F$_2$, the F--F distance.}
    \label{fig:precision}
\end{figure}

Figure~\ref{fig:precision} presents the bond dissociation curves and corresponding energy errors from the Full Configuration Interaction (FCI) baseline. 
For LiH (Figure~\ref{fig:precision}a), whose ground state is strongly single-reference near equilibrium, both the \emph{Hive} algorithm and the benchmarks effectively reproduce the potential energy surface. 
Across geometries, the robustness of the \emph{Hive} algorithm arises from combining chemistry-informed operator prioritisation during growth with a late-stage global amplitude rebalancing once the operator sequence stabilises.
This approach prevents the early misallocation of growth steps near equilibrium and eliminates parameter-coupling failures in the stretched regime.
To systematically analyse which components drive these gains in precision, we run a unified analysis of algorithmic mechanisms across bond lengths and metrics, which we discuss in detail in Section \ref{sec:interpretability}.

For H$_2$O, the challenge shifts to maintaining energy precision once symmetric O--H stretching drives the optimisation into a multi-reference regime.
In this regime, the adaptive baselines frequently plateau under their stopping criteria: after global reoptimisation, the best available single-operator extension no longer delivers a sufficient net improvement, leaving residual errors in the $\sim\qtyrange{e-2}{e-3}{\hartree}$ range at stretched geometries.
Across geometries, the \emph{Hive} algorithm maintains chemical precision by steering growth with geometry-aware operator prioritisation and staged search control, and—crucially for the stretched regime—by adding an explicit late-stage global parameter refinement over the stabilised operator sequence (see Appendix~\ref{app:ablation_study_h2o}).

For F$_2$ (Figure~\ref{fig:precision}c), the \emph{Hive} algorithm and the adaptive baselines reproduce the overall dissociation profile at chemical precision.
Here, the \emph{Hive} algorithm reaches chemical precision for F$_2$ by combining chemistry-informed operator ranking with growth-control logic that detects and repairs stalls in optimisation progress, and then applying a final global amplitude rebalancing once the ansatz stabilises (see Appendix~\ref{app:ablation_study_f2}).

Crucially, the distinction between bond lengths used in the evolution (hollow markers) and only during test time (solid markers) in Figure~\ref{fig:precision} demonstrates the generalisation power of the discovered algorithms across bond lengths.
In all three cases, the \emph{Hive} evolution was performed on a sparse subset of geometries but achieved high precision on unseen bond lengths.
This indicates that the discovered algorithms are not merely overfitting to specific training points but can generalise to unseen geometries.

\subsection{Reduction in Quantum Resources}
\label{sec:resources}
\begin{figure}[t]
    \centering
    \includegraphics{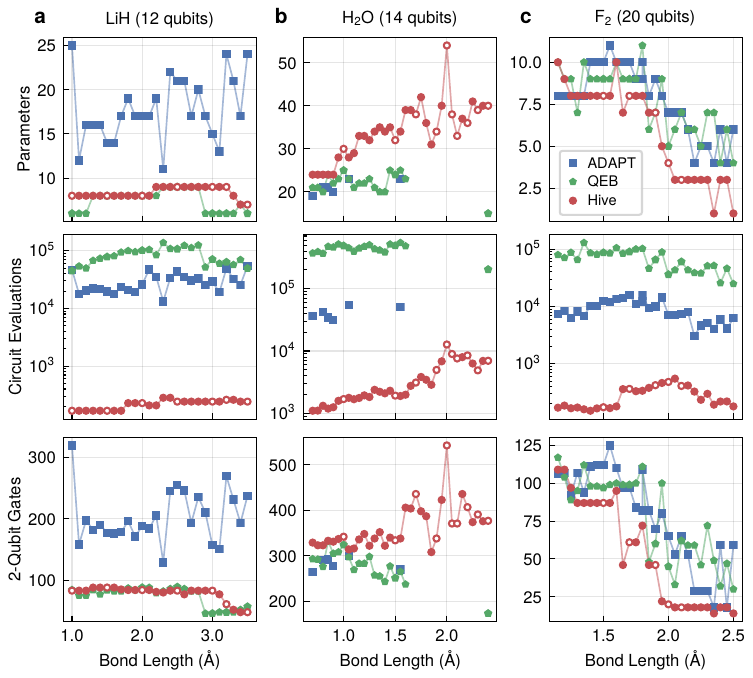}
    \caption{\textbf{Quantum resource overhead of the \emph{Hive} algorithms and ADAPT baselines for different molecules.} Scaling of variational parameters (top row), total circuit evaluations (exact statevector energy evaluations, middle row), and two-qubit gate counts (after compilation to the Quantinuum H2 native gate set, bottom row) for LiH (\textbf{a}), H$_2$O (\textbf{b}), and F$_2$ (\textbf{c}). \emph{Hive} results are shown as red circles, ADAPT-VQE and QEB-ADAPT-VQE are shown as blue squares and green pentagons, respectively. Points where algorithms failed to converge to energy error below chemical precision $E-E_{\mathrm{FCI}}\le 1.6\,\mathrm{mHa}$ are omitted from the plots. Bond lengths follow the definitions in Fig.~\ref{fig:precision}.}
    \label{fig:resources}
\end{figure}

We evaluate the quantum resource overhead required to reach chemical precision along dissociation scans of LiH, H$_2$O, and F$_2$ (Figure~\ref{fig:resources}), reporting ansatz size (number of variational parameters/operators), total circuit evaluations (exact statevector energy evaluations), and two-qubit gate counts after compilation to the Quantinuum H2 native gate set~\cite{_quantinuumquantinuumhardwarespecifications_2025}.
Although the \emph{Hive} algorithms, ADAPT-VQE, and QEB-ADAPT-VQE all grow a problem-tailored ansatz iteratively from the Hartree--Fock reference, their trajectories separate sharply once the chemical-precision filter is applied (points that fail $E-E_{\mathrm{FCI}}\le 1.6\,\mathrm{mHa}$ are omitted).

For LiH (Figure~\ref{fig:resources}a), the \emph{Hive} maintains a compact, single-digit operator sequence across the bond length range and reaches chemical precision with only a few hundred evaluations per geometry, whereas ADAPT-VQE and QEB-ADAPT-VQE require orders of magnitude more evaluations.
This ordering is consistent with the ADAPT-style outer loop in which each growth step estimates operator-pool gradients via Hamiltonian commutators and then re-optimises the accumulated ansatz parameters, and with QEB-ADAPT's added candidate-testing stage that trades additional evaluation overhead for circuit compactness.
Those performance gains are enabled by operator scoring as the key early mover that makes chemical precision reachable with a short ansatz, post-growth coordinate refinement which stabilises accuracy at stretched bonds without exploding the evaluation budget, and a final hardware-oriented discretisation step (angle snapping to a small discrete set) that removes small-amplitude structure while preserving the energy target; see Section~\ref{sec:interpretability}.

For H$_2$O (Figure~\ref{fig:resources}b), the chemical-precision filter exposes baseline coverage limits in the stretched regime, whereas the \emph{Hive}-evolved routine retains chemical precision throughout with tens of operators and a few hundred compiled two-qubit gates, while keeping evaluation overhead in the $10^3$--$10^4$ range—typically about an order of magnitude below ADAPT-VQE and about two orders below QEB-ADAPT-VQE where comparisons are available.
Here, the ablations (Appendix \ref{app:ablation_study_h2o}) indicate that H$_2$O's \emph{Hive} routine succeeds via geometry-adaptive pool shaping and singles-first growth to constrain the search, multi-angle local candidate testing to drive rapid in-growth improvement, and a brief coordinate-refinement-and-compress stage (gate-aware acceptance/pruning at large bond lengths) that preserves few-hundred-gate circuits.
Those strategies resemble an \enquote{engineer $\rightarrow$ test $\rightarrow$ refine} pattern aligned with pool-engineered adaptive-VQE resource reductions~\cite{ramoa_reducingresources_2025}, coordinate-wise gradient-free excitation optimizers~\cite{ostaszewski_structureoptimization_2021, jager_fastgradientfree_2025}, and orbital-optimized VQE strategies~\cite{zhao_orbitaloptimizedpaircorrelated_2023}.

For F$_2$, all three methods reach chemical precision with broadly similar ansatz sizes, but the \emph{Hive} collapses evaluation overhead to the few-hundred-call regime and becomes markedly leaner toward dissociation, often compressing to only a handful of operators and tens of compiled two-qubit gates.
The F$_2$ ablations (see Appendix \ref{app:ablation_study_f2}) attribute the chemical precision at stretched bonds to rejection-recovery local reoptimisation plus post-growth Newton/coordinate refinement, whereas the dissociation-regime gate compression is driven by explicit pruning of negligible operators during/after growth, consistent with pruning strategies proposed for adaptive ans\"atze and compilation-aware circuit simplification~\cite{vaquero-sabater_prunedadaptvqecompacting_2025, escofet_quantumcircuit_2026}.

\subsection{Algorithmic Mechanisms}
\label{sec:interpretability}
Evolutionary coding agents like the \emph{Hive} return an explicit algorithm, including high-level descriptors, allowing direct inspection of the evolved logic~\cite{romera-paredes_mathematicaldiscoveries_2024, novikov_alphaevolvecoding_2025, zhang_quantumcomputation_2025}, rather than encoding their strategy in learned parameters or the surrounding search procedure~\cite{krenn_scientificunderstanding_2022, wetzel_interpretablemachine_2025, rudin_stopexplaining_2019}.
Across molecules, the evolved solvers follow the same adaptive loop: starting from the Hartree--Fock state, they propose candidate excitations from the implemented operator pools, update parameters, accept operators only if they pass an improvement test, and terminate when the shared termination criteria are met.
We factor this loop into a mechanism ladder starting from a minimal baseline (L0): operator scoring (L1), growth control (L2), optimisation (L3), and post-processing that refines parameters (L4) and compresses the ansatz (L5). 
Activating these mechanisms sequentially isolates their impact on energy error and resource cost while keeping evolved implementation accelerations fixed (see Appendix~\ref{app:ablation_studies}).

In Figure~\ref{fig:ablation_LiH}, we deconstruct the evolved LiH algorithm and measure the effect of each level on energy error, circuit evaluations, and two-qubit gate count (Figure~\ref{fig:ablation_LiH}a–c), with the ladder and notation summarised in Figure~\ref{fig:ablation_LiH}d. 
Resource panels report only configurations within the chemical precision threshold; corresponding ablation studies for H$_2$O and F$_2$ are reported in Appendix~\ref{app:ablation_study_h2o} and \ref{app:ablation_study_f2}.
\begin{figure*}[p!]
    \centering
    \includegraphics{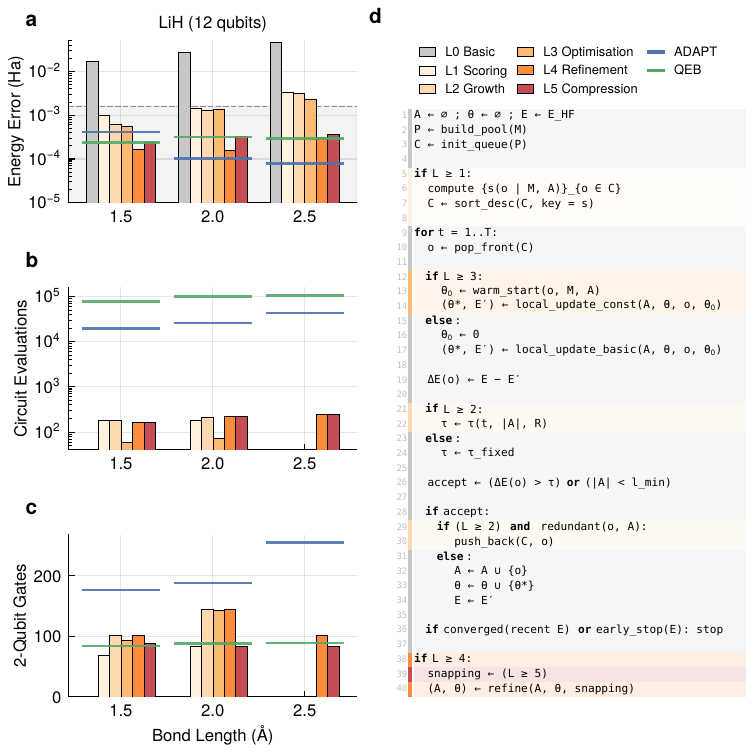}
    \caption{\textbf{LiH ablations and mechanism-level interpretability of the \emph{Hive}-evolved algorithm.} (\textbf{a--c}) Progressive ablation study for LiH (12 qubits) across bond lengths. Bars show the performance of the mechanism ladder (L0 Basic $\rightarrow$ L1 Scoring $\rightarrow$ L2 Growth $\rightarrow$ L3 Optimisation $\rightarrow$ L4 Refinement $\rightarrow$ L5 Compression) for energy error (\textbf{a}), circuit evaluations (\textbf{b}), and two-qubit gate count (\textbf{c}), with colours matching the ladder levels. Horizontal blue (green) lines denote ADAPT-VQE (QEB-ADAPT-VQE) results at the same geometries. The dashed horizontal line and shaded region indicate the chemical-precision band; resource panels report only configurations that meet the precision threshold $E-E_{\mathrm{FCI}}\le 1.6\,\mathrm{mHa}$. (\textbf{d}) Single pseudocode block showing the ablation level $L\in\{0,1,2,3,4,5\}$, where higher levels activate additional branches (colour-coded as in \textbf{a--c}). Symbols: $M$ molecular data at fixed geometry; $R$ bond length; $A$ current operator set (ansatz); $\theta$ current parameter vector; $E$ current energy; $P$ fixed operator pool; $C$ candidate queue; $s$ operator score; $o\in C$ candidate operator; $t$ growth-iteration index; $T$ maximum growth iterations; $\theta_0$ initial angle for the newly proposed operator; $\theta^\star$ updated angle returned by the local update; $E'$ trial energy after the local update; $\Delta E(o)=E-E'$ per-append energy improvement; $\tau$ acceptance threshold; $l_{\min}$ minimum enforced ansatz length.}
    \label{fig:ablation_LiH}
\end{figure*}

At level L0, we reduce the solver to a minimal baseline without an explicit mechanism for choosing promising excitations or coordinating parameters. 
It traverses a fixed candidate queue and, at each step, optimises only the newly introduced angle from a zero initialisation using a generic 1D local search.
The algorithm appends the candidate if the resulting energy decrease clears a fixed, geometry-independent threshold (or until a minimum ansatz length is enforced); otherwise, it discards it. 
As seen in Figure~\ref{fig:ablation_LiH}a, this baseline fails to reach chemical precision for LiH at the tested geometries and therefore does not enter the resource comparisons in Figure~\ref{fig:ablation_LiH}b--c.

Level L1 activates scoring to make the operator choice informative. 
Relative to L0's fixed order, it ranks candidate excitations using mean-field chemistry proxies together with a soft locality bias that down-weights large orbital span.
Such MP2-amplitude-guided heuristics closely parallel chemistry-motivated operator ranking schemes proposed for variational algorithms~\cite{romero_strategiesquantum_2018, haidar_opensource_2023, vaquero-sabater_physicallymotivated_2024, fedorov_unitaryselective_2022, majland_fermionicadaptive_2023, halder_efficientquantum_2025, li_efficientrobust_2024} and derive from the perturbative structure underlying MP2~\cite{moller_noteapproximation_1934, head-gordon_mp2energy_1988}.
Crucially, L1 changes only the ordering of candidate operators—the acceptance test and local update routine remain as in L0—thereby isolating the effect of improved operator prioritisation during growth.
In Figure~\ref{fig:ablation_LiH}a, L1 achieves the dominant near-equilibrium precision gain, reaching chemical precision at $\qty{1.5}{\angstrom}$ and $\qty{2.0}{\angstrom}$, but remains above threshold at $\qty{2.5}{\angstrom}$, consistent with the reduced reliability of single-reference MP2-like hierarchies under increased static correlation \cite{hollett_twofaces_2011, romero_strategiesquantum_2018}.

Level L2 activates growth control to regulate expansion beyond L1's ranking.
It replaces the fixed acceptance test with an adaptive, geometry- and iteration-dependent threshold that relaxes as the ansatz grows, and it defers near-duplicate excitations by re-queuing them.
This behaviour mirrors the tunable selection and stopping thresholds used in ADAPT-style growth~\cite{grimsley_adaptivevariational_2019, grimsley_adaptiveproblemtailored_2023}, and is consistent with evidence that adaptive ansätze can accumulate low-impact operators that can later be suppressed with minimal loss in energy~\cite{vaquero-sabater_prunedadaptvqecompacting_2025}. 
In Figure~\ref{fig:ablation_LiH}a, L2 adds only a modest precision gain over L1 near equilibrium and can increase evaluations and two-qubit gates by permitting larger ansätze (Figure~\ref{fig:ablation_LiH}b--c), while remaining above threshold at $\qty{2.5}{\angstrom}$.

Level L3 improves optimisation to reduce the per-append evaluation cost.
It replaces zero initialisation and a variable-length 1D search with a warm-started, constant-budget coordinate update that relies on the same mean-field ingredients used for scoring. 
This design is closely related to gradient-free coordinate-descent optimisers such as Rotosolve~\cite{ostaszewski_structureoptimization_2021} and ExcitationSolve~\cite{jager_fastgradientfree_2025}, which exploit the low-degree trigonometric structure of single-parameter circuit slices to obtain accurate 1D updates from few energy evaluations~\cite{li_efficientrobust_2024, feniou_greedygradientfree_2025}.
In Figure~\ref{fig:ablation_LiH}b, L3 yields the dominant reduction in circuit evaluations at essentially fixed precision and gate count (Figure~\ref{fig:ablation_LiH}a,c), but it remains above the chemical-precision threshold at $\qty{2.5}{\angstrom}$ (Figure~\ref{fig:ablation_LiH}a).

Level L4 activates post-refinement after growth.
With the operator set fixed, it runs one or more global refinement sweeps that re-optimise all angles using the same constant-budget 1D primitive as L3.
This step compensates for the one-at-a-time tuning during growth, which can leave earlier angles suboptimal—especially at stretched geometries where static correlation strengthens inter-parameter coupling \cite{hollett_twofaces_2011, grimsley_adaptivevariational_2019}.
Accordingly, L4 is the first level to reach chemical precision at $\qty{2.5}{\angstrom}$ and further improves the near-equilibrium points (Figure~\ref{fig:ablation_LiH}a). 
Its signature cost is a higher evaluation count (Figure~\ref{fig:ablation_LiH}b), while two-qubit gates remain essentially unchanged (Figure~\ref{fig:ablation_LiH}c).

Level L5 targets hardware cost after L4 has recovered an accurate solution.
During post-refinement, it snaps parameters to a small discrete set and prunes near-zero rotations, reducing the effective operator set.
This yields the dominant reduction in two-qubit gate count relative to L4 while preserving chemical-precision energies (Figure~\ref{fig:ablation_LiH}a,c) with negligible additional evaluation overhead (Figure~\ref{fig:ablation_LiH}b). This behaviour aligns with removing low-impact operators from adaptive ansätze \cite{vaquero-sabater_prunedadaptvqecompacting_2025} and with compilation-aware pruning of small-angle parametric gates \cite{escofet_quantumcircuit_2026}.

Overall, the mechanisms separate cleanly by function: scoring and global refinement drive energy gains, constant-budget local updates reduce circuit evaluations, and snapping reduces the two-qubit gate count without sacrificing chemical precision.

\subsection{Experimental Robustness and Hardware Implementation}
\begin{figure}[t]
    \centering
    \includegraphics{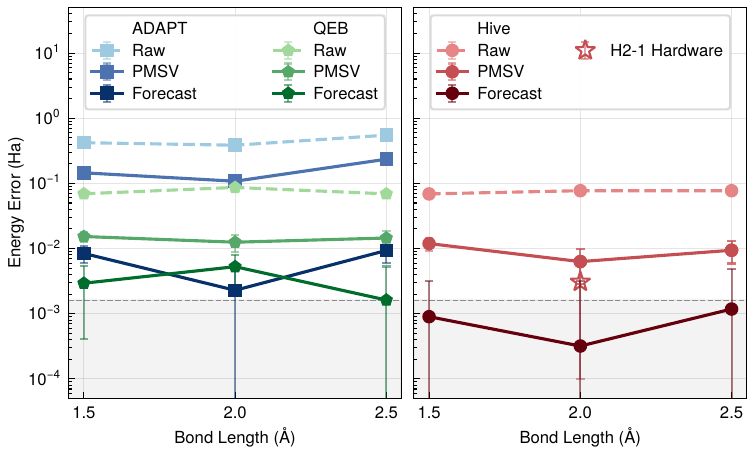}
    \caption{\textbf{Robustness and error mitigation effectiveness of the LiH \emph{Hive} algorithm and (QEB-)ADAPT-VQE on noisy emulator and hardware.} The left panel displays results for ADAPT-VQE (blue squares) and QEB-ADAPT-VQE (green pentagons), while the right panel shows the \emph{Hive} algorithm (red circles). Results are presented for three noise regimes: unmitigated emulator results (Raw, dashed lines), mitigated results using Partition Measurement Symmetry Verification (PMSV, lighter solid lines), and a Forecast model (darker solid lines). Additionally, we present experimental data obtained from the Quantinuum H2-1 hardware (red star at $\qty{2.0}{\angstrom}$). The Forecast model utilizes Quantinuum H2-1 Emulator parameters \cite{_quantinuumquantinuumhardwarespecifications_2025} with single-qubit error rates tuned from $10^{-5}$ to $10^{-7}$ and two-qubit error rate tuned from $10^{-3}$ to $10^{-5}$. Memory errors in this model are improved by a factor of 2--4. The horizontal dashed grey line marks the chemical precision threshold. All points use a fixed shot budget per measurement circuit of $10^4$ shots, and error bars are obtained by bootstrap resampling. Bootstrap error bars appear geometrically asymmetric on logarithmic scales due to the non-linear mapping of the y-axis.}
    \label{fig:emulator}
\end{figure}

We benchmark ADAPT-VQE, QEB-ADAPT-VQE, and \emph{Hive}-generated ansätze for LiH at $R =\qtylist{1.5;2.0;2.5}{\angstrom}$ on the Quantinuum H2-1 emulator~\cite{_quantinuumquantinuumhardwarespecifications_2025} and report the deviation from the FCI reference energy (Figure~\ref{fig:emulator}).
We evaluate three noise regimes: (i) the default emulator noise model (\enquote{Raw}), (ii) the same model using the Partition-Measurement Symmetry Verification (PMSV) \cite{yamamoto_quantumhardware_2022} noise mitigation of \emph{InQuanto}, and (iii) a tuned \enquote{Forecast} noise model intended to reflect improved device parameters.

Under the default emulator noise model, PMSV reduces the energy bias for all methods, with the largest reduction observed for the \emph{Hive} circuits.
With PMSV enabled, \emph{Hive} remains systematically closest to FCI, achieving $\Delta E_{\mathrm{FCI}}\approx 6$--$12$ mHa across the three geometries, compared with $\approx 13$--$15$ mHa for QEB-ADAPT-VQE and $\gtrsim 10^2$ mHa for ADAPT-VQE.
For LiH at $R=\qty{2.0}{\angstrom}$, we additionally execute the \emph{Hive}-discovered circuit on the Quantinuum System Model H2-1 quantum computer \cite{quantinuumh21runs}. 
With PMSV enabled, the measured energy deviates from the FCI reference by $3.1\,\mathrm{mHa}$ with a statistical uncertainty of $3.4\,\mathrm{mHa}$, making this hardware result statistically consistent with chemical precision at this geometry within the reported uncertainty interval.
Among prior LiH ground-state calculations on quantum hardware, as surveyed in Ref.~\cite{yoffe_qubitefficientvariational_2025}, demonstrations typically rely on strongly reduced encodings involving only a few qubits.
In contrast, our H2-1 experiment targets the full 12-qubit LiH encoding used throughout this work, providing an end-to-end validation that the \emph{Hive}-generated ansatz maintains mHa-level agreement with FCI under native device noise.

To probe performance under improved yet still realistic conditions, we also consider a tuned Forecast noise model with reduced single- and two-qubit error rates and improved memory errors (see Figure~\ref{fig:emulator} caption).
Under the tuned Forecast model, the \emph{Hive} circuit reaches chemical precision across all three geometries ($\Delta E_{\mathrm{FCI}}\approx 0.3$--$1.2~\mathrm{mHa}$), whereas QEB-ADAPT-VQE and ADAPT-VQE remain above chemical precision at most geometries, indicating that the \emph{Hive} algorithm's precision advantage persists under realistic noise and strengthens under improved hardware conditions.

\section{Discussion}
This work demonstrates that autonomous, AI-driven program search can discover quantum routines for ground-state estimation that outperform established human-designed methods. 
In particular, the functions produced by the \emph{Hive} for algorithmic discovery achieve higher precision than adaptive baselines on the tested molecular systems, with orders-of-magnitude fewer circuit evaluations and comparable or fewer two-qubit gates. 
In the NISQ setting, where sampling cost and noise limit the practical depth and width of circuits, such gains are a strong indicator that AI-discovered heuristics can meaningfully complement existing strategies.

A central feature of the present approach is that the \emph{Hive} does not construct quantum circuits instance-by-instance. Instead, it discovers and refines a classical function, \texttt{generate\_ansatz()}, that maps problem data and operator pools to a full circuit specification. Once discovered, this function can be reused across bond lengths or related problem instances to generate circuits in a consistent, algorithmic manner. This generative view distinguishes the method from one-off circuit optimisation: the outcome of the search is an explicit, human-readable routine that encodes heuristics for operator selection, ordering, truncation, and parameter handling, and can be inspected, benchmarked, or further adapted by practitioners.

The flexibility of the framework suggests a broader applicability beyond the specific electronic-structure benchmarks studied here. Because the optimisation target is defined at the level of classical evaluation logic, one can, in principle, direct the search toward alternative metrics, such as gate depth, two-qubit gate count, or robustness to noise, or toward hybrid objectives that balance precision and resource cost. Naturally, the same infrastructure could be applied to other near-term quantum algorithms in areas such as combinatorial optimisation, provided that a suitable evaluation loop and operator pool are specified.

The methods presented here also invite exploration in the context of fault-tolerant quantum computation. For example, early demonstrations of error-corrected quantum phase estimation for quantum chemistry~\cite{yamamoto_quantumerrorcorrected_2025} already show that implementations of nontrivial algorithms using logically encoded qubits and operations are becoming feasible. In such regimes, AI-discovered state-preparation or ansatz-generation routines could be optimised, for example, by minimising logical depth, thereby tailoring algorithmic structure to the constraints of specific architectures.

Despite these promising results, limitations should be acknowledged. First, the discovery process itself is computationally intensive. It requires repeated evaluations of quantum circuits, which, in case they are simulated classically, currently constrain the size of the systems and operator pools that can be explored.
Such upfront cost is viable in the case where the discovered quantum algorithm showcases good generalisation on other problem instances, the full study of which is left for future work.
Second, the search space is still defined by human choices such as the operator pool, circuit representation, and prompt design. These design decisions may bias the class of algorithms that can be found. At the same time, this can be a feature allowing a level of control over the direction of the evolution.

Looking ahead, the increasing performance and availability of high-level quantum software frameworks, including those used in this work, provide a growing platform for integrating AI-assisted algorithm discovery across the quantum stack. Beyond ansatz construction for chemistry, similar techniques could target compilation passes, layout and routing strategies on quantum chips, low-level control sequences, or classical algorithms relevant to quantum error correction, such as decoders. Treating quantum algorithms as programs to be written, refactored, and optimised by AI systems, rather than as fixed circuit templates, may be a productive paradigm for both NISQ devices and emerging fault-tolerant platforms.

~\\
{\bf Acknowledgements:} We thank Abbey Pint for figure design, Eric Brunner for project management, Simon McAdams and Andrew Tranter for \emph{InQuanto} access, Ciaran Ward for advice on emulator runs, Dan Gresh, Peter Siegfried, and Joshua Savory for assistance with execution on Quantinuum's H2-1 quantum processor, and Duncan Gowland, Kripa Panchagnula and Luca Erhart for advice on error mitigation experiments.

\bibliography{Hiverge}

\newpage

\appendix

\section{Problem Description}
\subsection{Hamiltonian}
\label{app:hamiltonian}
Under the Born--Oppenheimer approximation, the nuclei are treated as fixed classical point charges.
For a fixed nuclear configuration, the electronic problem is therefore described by~\cite{mcardle_quantumcomputational_2020, szabo_modernquantum_2012}
\begin{equation}
H =
-\sum_i \frac{\nabla_i^2}{2}
-\sum_{i,I}\frac{Z_I}{|\mathbf{r}_i-\mathbf{R}_I|}
+\frac{1}{2}\sum_{i\ne j}\frac{1}{|\mathbf{r}_i-\mathbf{r}_j|}.
\end{equation}
Here $i$ and $j$ index electrons, $I$ indexes nuclei, $Z_I$ is the atomic number of nucleus $I$, $\mathbf{r}_i$ is the position of electron $i$, $\mathbf{R}_I$ is the position of nucleus $I$, and $\nabla_i^2$ is the Laplacian with respect to the coordinates of electron $i$.
The three terms correspond to the electron kinetic energy, electron--nucleus attraction, and electron--electron Coulomb repulsion in atomic units, where the unit of length is $a_0 = \qty{0.529e-10}{\meter}$, the unit of mass is the electron mass $m_e$, and the unit of energy is 1 Ha.

To obtain a finite representation, we project the Hamiltonian onto $M$ basis wavefunctions $\{\phi_p(x_i)\}$, where $x_i=(\mathbf{r}_i,\sigma_i)$ with $\sigma_i$ the spin coordinate of electron $i$.
In this basis, the Hamiltonian takes the second-quantised form~\cite{mcardle_quantumcomputational_2020, szabo_modernquantum_2012}
\begin{equation}
H=
\sum_{p,q} h_{pq} a_p^\dagger a_q
+\frac{1}{2}\sum_{p,q,r,s} h_{pqrs} a_p^\dagger a_q^\dagger a_r a_s.
\end{equation}
Here $p,q,r,s$ label spin orbitals, $a_p^\dagger$ and $a_p$ are fermionic creation and annihilation operators acting on spin orbital $p$, and $h_{pq}$ and $h_{pqrs}$ are the one- and two-electron integrals in the chosen basis.
For the computations reported here, molecular Hamiltonians are generated using Quantinuum's \emph{InQuanto}~\cite{inquanto} together with \textit{PySCF}~\cite{sun_recentdevelopments_2020}.
We use a Restricted Open Shell Hartree--Fock (ROHF) reference in the STO-3G basis~\cite{mcardle_quantumcomputational_2020, szabo_modernquantum_2012}.
The resulting fermionic Hamiltonian is then mapped to qubits as described in the following Section.

\subsection{Qubit Encoding and Excitation Operators}
\label{app:qubit_excitations}
To represent the second-quantised Hamiltonian on qubits, we use the Jordan--Wigner transformation, which stores the occupation of spin orbital $p$ in qubit $p$ and maps fermionic operators to Pauli strings~\cite{jordan_uberpaulische_1928, mcardle_quantumcomputational_2020}.
The resulting qubit Hamiltonian is therefore a sum of Pauli strings.

The operator pools used in this work are built from particle-number-preserving qubit excitation operators~\cite{yordanov_efficientquantum_2020, arrazola_universalquantum_2022}.
Their orbital structure is motivated by UCCGSD and by the more compact $k$-UpCCGSD family~\cite{anand_quantumcomputing_2022, lee_generalizedunitary_2019}.
Rather than exponentiating Jordan--Wigner-mapped fermionic excitations directly, these excitations are implemented as direct qubit excitation gates~\cite{yordanov_efficientquantum_2020, arrazola_universalquantum_2022}.
Single excitations are realised as Givens-type $U(2)$ rotations on the $\{|01\rangle,|10\rangle\}$ subspace, while the corresponding double-excitation gates act as $U(2)$ rotations on the $\{|0011\rangle,|1100\rangle\}$ subspace and leave the remaining basis states unchanged~\cite{arrazola_universalquantum_2022}.
This yields a compact, particle-number-preserving operator pool suitable for the variational ans\"atze studied in this work~\cite{yordanov_efficientquantum_2020, yordanov_qubitexcitationbasedadaptive_2021}.

\subsection{Baselines}
\label{app:baselines}
We benchmark the \emph{Hive}-discovered algorithms against ADAPT-VQE~\cite{grimsley_adaptivevariational_2019} and QEB-ADAPT-VQE~\cite{yordanov_qubitexcitationbasedadaptive_2021} using the same molecule-specific qubit-excitation pool families~\cite{yordanov_efficientquantum_2020} identified from the corresponding \emph{Hive} solvers, so that the comparison isolates differences in adaptive growth and optimisation rather than pool expressivity. 

In both baseline implementations, the variational parameters are optimised using SciPy's L-BFGS-B algorithm with analytic gradients, and the gradients of the generalised single- and double-excitation gates are evaluated via a generalised four-term parameter-shift rule~\cite{wierichs_generalparametershift_2022, arrazola_universalquantum_2022}.
ADAPT-VQE computes the full pool-gradient vector, appends the operator with the largest absolute gradient initialised at zero, and then fully reoptimises the accumulated ansatz. QEB-ADAPT-VQE instead uses the same gradient ranking only to define a shortlist, fully reoptimises the top-$k$ candidates with the newly added parameter initialised at zero, and appends the candidate that yields the largest realised energy decrease; in all reported QEB runs, we set $k=10$. 
In the implementation, ADAPT monitors three stopping checks---maximum gradient magnitude~\cite{grimsley_adaptiveproblemtailored_2023}, step-to-step energy improvement, and vanishing newest parameter~\cite{vaquero-sabater_prunedadaptvqecompacting_2025}---whereas QEB terminates when the best look-ahead energy decrease falls below a preset threshold. 
In the reported benchmark outputs, ADAPT exits almost exclusively through the energy-improvement criterion, with a single F$_2$ point stopping on the vanishing-parameter criterion, while all QEB runs terminate through the look-ahead threshold. For QEB, this minimum realized improvement is set to $10^{-4}$ Ha for LiH and $10^{-6}$ Ha for H$_2$O and F$_2$. 

\section{Ablation studies of discovered codes}
\label{app:ablation_studies}
In Section~\ref{sec:interpretability} and throughout this appendix, we analyse the final molecule-specific solvers through bottom-up, cumulative ablations. 
For each molecule, we start from a minimal baseline inherited from the full solver and then enable additional mechanisms sequentially in an L0-L5 ladder. 
The level labels are shared across LiH, H$_2$O, and F$_2$, but the concrete routines assigned to each level are molecule-specific, so the ladder should be read as a common functional decomposition rather than as identical code blocks across molecules. 
For each molecule and each tested bond length, we hold fixed the underlying electronic-structure instance—the Hamiltonian at that geometry, the basis/orbital-space choice, and the Hartree–Fock reference state—as well as the operator encodings, the backend energy-evaluation pathway, and the implementation accelerations inherited from the full solver, including energy caching, molecular-data caching, and reuse of parametrised circuit kernels. 
Shared stopping infrastructure is likewise retained; where the evolved code uses mode-dependent schedules, we treat those schedules as part of growth control and enable them only from L2 onward. 
Accordingly, L0 should be read as a reduced inherited baseline rather than a from-scratch null model: it retains the shared solver infrastructure while disabling the higher-level decision logic introduced at later levels.

This decomposition is an interpretability scaffold rather than a unique parsing of the code. 
We assign routines to levels by manual code inspection and LLM assistance, used only to compare and organise recurring motifs across the final evolved codes. Some helper routines could reasonably be placed one level earlier or later, but we use the ordering that most clearly separates operator scoring, growth control, optimisation, post-growth refinement, and compression. 
The appendix, therefore, analyses the final discovered codes themselves, rather than the full evolutionary trajectory or prompt history.

\subsection{H\texorpdfstring{$_2$}{2}O code}
\label{app:ablation_study_h2o}
\begin{figure*}[p!]
    \centering
    \includegraphics{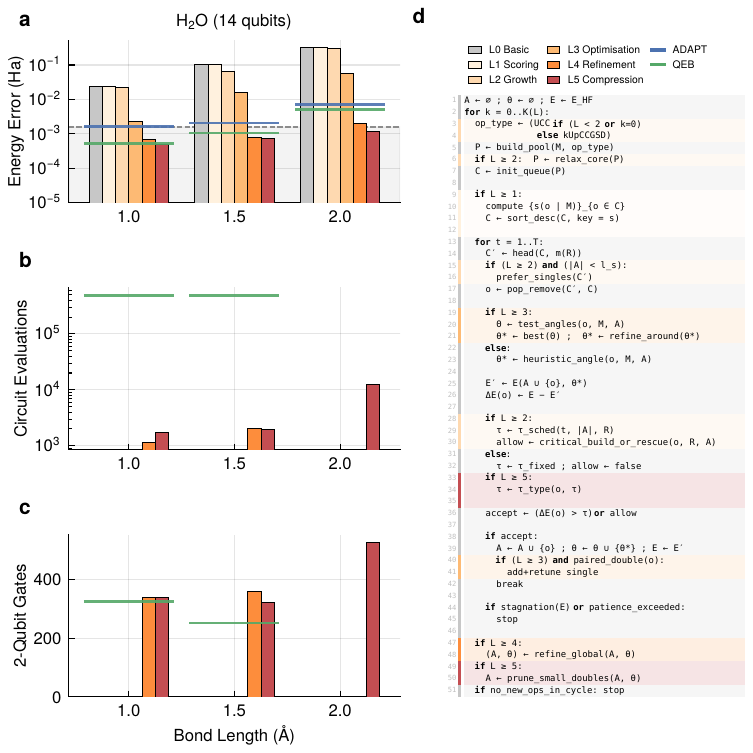}
    \caption{\textbf{H$_2$O ablations and mechanism-level interpretability of the \emph{Hive}-evolved algorithm.} (\textbf{a--c}) Progressive ablation study for H$_2$O (14 qubits) across bond lengths. Bars show the performance of the mechanism ladder (L0 Basic $\rightarrow$ L1 Scoring $\rightarrow$ L2 Growth $\rightarrow$ L3 Optimisation $\rightarrow$ L4 Refinement $\rightarrow$ L5 Compression) for energy error (\textbf{a}), circuit evaluations (\textbf{b}), and two-qubit gate count (\textbf{c}), with colours matching the ladder levels. Horizontal blue (green) lines denote ADAPT-VQE (QEB-ADAPT-VQE) results at the same geometries. The dashed horizontal line and shaded region indicate the chemical-precision band; resource panels report only configurations that meet the precision threshold $E-E_{\mathrm{FCI}}\le 1.6\,\mathrm{mHa}$. (\textbf{d}) Single pseudocode block showing the ablation level $L\in\{0,1,2,3,4,5\}$, where higher levels activate additional branches (colour-coded as in \textbf{a--c}). Symbols: $M$ molecular data at fixed geometry; $R$ bond length; $k$ outer self-training-cycle index; $K(L)$ maximum number of outer training cycles; $\mathrm{op\_type}$ operator-pool family used in the current outer cycle; $P$ operator pool for the current outer cycle; $C$ candidate queue; $C'$ temporary working head slice used in the current selection step; $s$ operator score; $m(R)$ bond-length-dependent head-slice size; $A$ current operator set (ansatz); $|A|$ current ansatz length; $l_s$ singles-first warmup cutoff; $o \in C$ candidate operator; $\Theta$ test-angle set for the proposed operator; $\theta$ current parameter vector; $\theta^\star$ selected/refined angle for the newly proposed operator; $E$ current energy; $E'$ trial energy after appending $o$ at $\theta^\star$; $\Delta E(o)=E-E'$ per-append energy improvement; $\tau$ acceptance threshold; $t$ per-cycle growth-iteration index; $T$ maximum selection iterations.}
    \label{fig:ablation_h2o}
\end{figure*}

We analyse the evolved H$_2$O algorithm using the same mechanism ladder as Section~\ref{sec:interpretability} and measure the effect of each level on energy error, circuit evaluations, and two-qubit gate count (Figure~\ref{fig:ablation_h2o}a–c), with the ladder and notation summarised in Figure~\ref{fig:ablation_h2o}. 
Resource panels report only configurations within the chemical-precision threshold.
At level L0, the H$_2$O solver is reduced to a bare append-and-test skeleton. 
Starting from an empty ansatz, it builds a single baseline operator pool, forms a candidate queue, takes proposed excitations from the front of that queue, assigns each one a single heuristic trial angle—MP2-like for doubles, and a small bond-length-dependent fallback for singles—evaluates the trial energy after appending it, and accepts the update only when the resulting energy improvement clears the fixed baseline threshold.
In this form, the routine remains well above chemical precision at all three tested bond lengths and therefore does not enter the resource comparisons in Figure~\ref{fig:ablation_h2o}b--c.

At level L1, the H$_2$O solver activates the score-and-sort branch of the loop. Starting from the same baseline pool, it computes a score for each excitation and reorders the candidate queue accordingly. 
In that scoring step, double excitations with vanishing MP2-like two-electron-integral coupling are scored as zero, and outside the $\qtyrange{1.9}{2.2}{\angstrom}$ critical region, the code likewise scores weak non-paired doubles as zero once the bond length reaches $\qty{2.0}{\angstrom}$. 
Single excitations are ranked primarily by inverse-gap criteria, favouring orbital-rotation moves, while at stretched geometries the scoring additionally boosts paired doubles and doubles with larger orbital overlap. 
The growth loop itself, however, is unchanged: candidates are still taken from the front of the reordered queue, tested at a single heuristic trial angle, and accepted only when the resulting energy improvement clears the same fixed baseline threshold. 
In this form, L1 changes only which operators are tried first, not how they are locally tested or admitted into the ansatz. For H$_2$O, that reordering alone has almost no effect on accuracy, so the routine still does not reach chemical precision at all three tested bond lengths.

At level L2, the H$_2$O solver activates growth control, so the score-and-sort queue is no longer used inside a single fixed baseline pass. 
Instead, the pool can be rebuilt over up to four outer self-training cycles, using a UCC pool in the first cycle and single-layer $k$-UpCCGSD thereafter, while a relaxed frozen-core filter removes only pure core-core excitations and retains core--valence mixing moves. 
Within each cycle, growth becomes bond-length aware: the prioritised pool expands at stretched geometries, the convergence threshold and the number of no-improvement steps allowed before stopping are adjusted with bond length, early growth is biased toward singles while the ansatz is still short, the $\qtyrange{1.9}{2.2}{\angstrom}$ region is treated more leniently, and at larger stretch a dedicated paired-double building phase can admit important paired doubles early. 
The local test, however, remains unchanged: each candidate is evaluated at a single heuristic trial angle rather than through a structured angle scan. 
In this form, L2 broadens and stabilises the search, but it still does not recover sufficient correlation to reach chemical precision and therefore does not enter the resource comparisons in Figure~\ref{fig:ablation_h2o}b--c.

At level L3, the H$_2$O solver activates structured local optimisation during growth. 
The L2 growth-control scaffold remains in place, but candidate admission is no longer governed by a one-shot heuristic-angle test.
The algorithm now builds a bond-length-dependent test-angle set for that candidate, selects the best local angle from that set, and then refines around that value on a denser non-uniform local grid. 
When a paired double excitation is accepted, the code can additionally insert the matching single excitation and retune that single locally. 
In this form, L3 is the first stage that produces a major reduction in energy error for H$_2$O, indicating that accurate append-time angle selection becomes essential once the problem enters the harder stretched-bond regime. 
That gain, however, comes with a pronounced increase in circuit evaluations and two-qubit gates, and the routine still remains above chemical precision at all three tested bond lengths, so it does not yet enter the resource comparisons in Figure~\ref{fig:ablation_h2o}b--c.

At level L4, the H$_2$O solver activates global post-growth refinement at fixed ansatz structure. 
After the L3 growth loop terminates, it runs a refinement sweep over the assembled ansatz, re-optimising all angles in a decoupled order that treats single excitations first and double excitations second, using coordinate-wise parabolic updates; for stretched geometries, and especially in the $\qtyrange{1.9}{2.2}{\angstrom}$ critical region, the code performs additional refinement passes. 
In contrast to L3, this stage no longer changes which operators are present in the ansatz, but instead rebalances the amplitudes of the already-grown ansatz. 
In this form, L4 is the primary cleanup stage for H$_2$O: it sharply reduces the residual energy error while also reducing the compiled two-qubit gate count relative to the L3 ansatz. 
It is the first level that reaches chemical precision at the shorter two bond lengths and therefore enters the resource comparisons in Figure~\ref{fig:ablation_h2o}b--c there, although the $\qty{2.0}{\angstrom}$ case still remains slightly above threshold.

At level L5, the H$_2$O solver activates the full gate-aware cleanup step. During growth, it no longer uses the L2--L4 acceptance rule unchanged, but adjusts the threshold according to operator type: double excitations are admitted more conservatively, whereas single excitations are treated more leniently, especially in the early singles-first phase. After the L4 global refinement, the code then prunes weak double excitations with small amplitudes and reevaluates the energy of the reduced ansatz. 
In this form, L5 no longer only rebalances the existing ansatz, but also applies a final cost-aware selection and cleanup pass over that ansatz. 
For H$_2$O, this last step does not act as a uniform gate-compression stage across all bond lengths; instead, its main effect is to secure the stretched-bond endpoint, bringing the $\qty{2.0}{\angstrom}$ case below chemical precision and thereby allowing the full algorithm to enter the resource comparisons in Figure~\ref{fig:ablation_h2o}b--c at all three geometries.

\subsection{F\texorpdfstring{$_2$}{2} code}
\label{app:ablation_study_f2}
\begin{figure*}[ht!]
    \centering
    \includegraphics{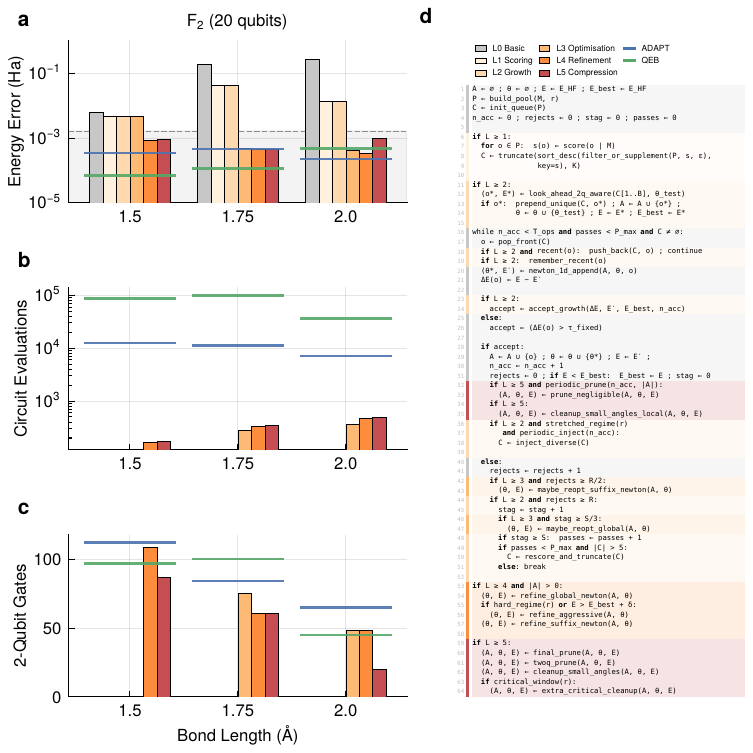}
    \caption{\textbf{F$_2$ ablations and mechanism-level interpretability of the \emph{Hive}-evolved algorithm.} (\textbf{a--c}) Progressive ablation study for F$_2$ (20 qubits) across bond lengths. Bars show the performance of the mechanism ladder (L0 Basic $\rightarrow$ L1 Scoring $\rightarrow$ L2 Growth $\rightarrow$ L3 Optimisation $\rightarrow$ L4 Refinement $\rightarrow$ L5 Compression) for energy error (\textbf{a}), circuit evaluations (\textbf{b}), and two-qubit gate count (\textbf{c}), with colours matching the ladder levels. Horizontal blue (green) lines denote ADAPT-VQE (QEB-ADAPT-VQE) results at the same geometries. The dashed horizontal line and shaded region indicate the chemical-precision band; resource panels report only configurations that meet the precision threshold $E-E_{\mathrm{FCI}}\le 1.6\,\mathrm{mHa}$. (\textbf{d}) Single pseudocode block showing the ablation level $L\in\{0,1,2,3,4,5\}$, where higher levels activate additional branches (colour-coded as in \textbf{a--c}). Symbols: $M$ molecular data at fixed geometry; $r$ bond length; $A$ current operator set (ansatz); $|A|$ current ansatz size; $\theta$ current parameter vector; $E$ current energy; $E_{\mathrm{best}}$ best energy seen so far; $P$ operator pool; $C$ candidate queue; $s(o)$ operator score; $o$ candidate operator; $o^\star$ look-ahead-selected operator; $B$ look-ahead batch size; $\theta_{\text{test}}$ look-ahead test angle; $E^\star$ energy after look-ahead seeding; $\theta^\star$ updated angle returned by the 1D local append update; $E'$ trial energy after appending $o$; $\Delta E(o)=E-E'$ per-append energy improvement; $\varepsilon$ score cutoff; $K$ queue truncation size; $n_{\mathrm{acc}}$ number of accepted operators; $T_{\mathrm{ops}}$ maximum accepted-operator budget; passes current pass count; $P_{\max}$ maximum number of passes.}
    \label{fig:ablation_f2}
\end{figure*}

Similarly, we analyse the evolved F$_2$ algorithm using the same mechanism ladder as Section~\ref{sec:interpretability} and measure the effect of each level on energy error, circuit evaluations, and two-qubit gate count (Figure~\ref{fig:ablation_f2}a--c), with the ladder and notation summarised in Figure~\ref{fig:ablation_f2}.
Resource panels report only configurations within the chemical-precision threshold.
For F$_2$, L0 isolates the minimal ansatz-growth loop: the algorithm builds a bond-length-dependent operator pool and places the resulting excitations in a simple candidate queue: at shorter bond lengths it uses a UCC pool, whereas in the stretched regime it switches to a UCCG pool.
It then traverses that queue in raw order and, for each proposed excitation, optimises only the newly appended angle with a single one-dimensional Newton step: it probes the trial ansatz at small displacements around zero, estimates the local gradient and curvature, clips the resulting update, and evaluates the trial energy after appending the candidate excitation.
The operator is accepted only if this per-append energy decrease exceeds the fixed baseline threshold. 
In this form, the routine contains no chemically informed ranking or adaptive search control: scoring and filtering are disabled, there is no look-ahead seeding, no recent-operator deferral, no multi-criterion acceptance, no stagnation-triggered repair, and no post-growth refinement or pruning. 
This L0 baseline, therefore, isolates the effect of the later F2-specific mechanisms from a minimal growth loop, and it remains above chemical precision at all three tested bond lengths, especially in the stretched regime.

At level L1, the F$_2$ solver makes operator choice informative before growth begins.
Relative to L0's raw queue order it assigns each pool excitation a chemistry-informed score, reorders the queue accordingly, filters out very weak candidates, and truncates the list to a bond-length-dependent size, supplementing from the top-ranked pool when needed. 
The score combines orbital energy and integral information with explicit orbital analysis features and hardware cost penalties, and it also applies spatial symmetry filtering when orbital symmetries are available. 
Crucially, L1 changes only how candidates are prioritised: the one-dimensional Newton append step and the fixed-threshold acceptance rule remain as in L0.
In this form, L1 provides the first meaningful improvement for F$_2$, especially at stretched bond lengths, but it still remains above chemical precision at all three tested geometries.

At level L2, the F$_2$ solver activates growth control, so the score-ranked queue is no longer consumed under a single fixed baseline rule. 
Before the main loop, it performs a short look-ahead over the top-scoring candidates and seeds the ansatz with the most promising one, akin to the QEB-ADAPT-VQE~\cite{yordanov_qubitexcitationbasedadaptive_2021} selection strategy. 
During growth, it keeps a short memory of recently used operators and defers immediate repeats, while candidate admission is no longer governed by the fixed L0/L1 threshold alone: instead, the code uses a multi-criterion acceptance rule that combines significant and progressive energy improvement with a more lenient early-growth condition, together with a two-qubit-gate efficiency check once the ansatz has started to grow. 
When progress stalls, the routine tracks rejections and passes, re-scores and truncates the remaining queue, and at stretched bond lengths can occasionally inject a deeper high-scoring excitation to diversify the search. 
The local append-time test itself, however, is unchanged: each candidate is still evaluated through the same single one-dimensional Newton append step used at L0 and L1. 
In this form, L2 makes the F$_2$ growth process more robust and better directed, but it still does not recover chemical precision at the three tested bond lengths.

At level L3, the F$_2$ solver adds stall-triggered local repair inside the growth loop. 
The L2 growth-control scaffold remains in place, but repeated rejections no longer simply accumulate toward a new pass or termination. 
Instead, once the search begins to stall, the code first reoptimises the most recently added subset of angles with a short Newton refinement and, if stagnation persists, then reoptimises the current ansatz more broadly before deciding whether to continue. 
If either repair lowers the energy, the stagnation counters are reset, and growth resumes from the repaired ansatz. 
In this form, L3 is the first level that recovers chemical precision at the stretched bond lengths, showing that for F$_2$ it is not sufficient to rank and admit promising excitations alone; the current ansatz must also be repaired once growth starts to stall. 
The shorter bond length, however, still remains above chemical precision, so L3 does not yet deliver target-precision performance uniformly across all three geometries.

At level L4, the F$_2$ solver activates global post-growth refinement at a fixed ansatz. 
Once the L3 growth loop terminates, the code no longer changes which operators are present in the ansatz; instead, it revisits the assembled operator sequence and reoptimises its parameters through a full Newton sweep, an additional aggressive pass for harder geometries or when the energy remains above the best-so-far target, and a final suffix Newton refinement. 
In contrast to L3, this stage no longer serves to repair stalled growth, but to rebalance the parameters of an already assembled ansatz. For F$_2$, L4 is therefore the main cleanup stage for the shorter bond length: it is the first level that brings the $\qty{1.5}{\angstrom}$ case to chemical precision, while the stretched-bond cases were already below threshold at L3 and are only refined further here. 
As a result, L4 is the first level that reaches chemical precision uniformly across all three tested geometries.

At level L5, the F$_2$ solver activates the full gate-aware compression stage. 
During growth, accepted updates are no longer left untouched: once the ansatz has become nontrivial, the code periodically prunes negligible operators and removes near-zero amplitudes when doing so does not materially worsen the energy. 
After the L4 post-growth refinement, it then applies a dedicated sequence of compression passes—a strong final pruning step, a two-qubit-gate-aware pruning pass, and a final small-angle cleanup—with more conservative thresholds in the 1.95--2.15~\AA\ critical window and an additional cleanup step there when needed. 
In contrast to L4, this stage is no longer primarily about recovering missing accuracy, but about stripping low-impact structure from an already accurate ansatz. 
For F$_2$, L5 is therefore the main compaction stage: unlike the LiH solver, it does not rely on angle snapping, but reduces the L4-refined ansatz through explicit pruning and gate-aware cleanup, which is exactly the mechanism that drives the leaner dissociation-regime circuits.

\end{document}